\documentclass[twocolumn]{aastex61}

\usepackage{amsmath,amstext}
\usepackage[T1]{fontenc}
\usepackage[figure,figure*]{hypcap}

\received{\today}
\revised{\today}
\accepted{\today}
\submitjournal{ApJ}

\shorttitle{GRAVITY/VLTI study of HD~93\,206~A}
\shortauthors{J. Sanchez-Bermudez et al.}

\begin{document}

\title{GRAVITY spectro-interferometric study of the massive multiple
  stellar system HD~93\,206~A}

\correspondingauthor{Joel Sanchez-Bermudez}
\email{jsanchez@mpia.de}

\author{J. Sanchez-Bermudez}
\affiliation{Max-Planck-Institut f\"ur Astronomie, K\"onigstuhl 17,
     Heidelberg 69\,117, Germany}
\author{A. Alberdi}
\affiliation{Instituto de Astrof\'{\i}sica de Andaluc\'{\i}a (CSIC), Glorieta de la Astronom\'{\i}a S/N, 18\,008 Granada, Spain}
\author{R. Barb\'a}
\affiliation{Departamento de F\'{\i}sica, Universidad de la Serena,
        Av. Cisternas 1200 Norte, 204\,000 La Serena, Chile}
\author{ J. M. Bestenlehner}
\affiliation{Max-Planck-Institut f\"ur Astronomie, K\"onigstuhl 17,
     Heidelberg 69\,117, Germany}
\affiliation{Department of Physics and Astronomy, University of Sheffield, Hicks Building, Hounsfield Rd, Sheffield, S3 7RH, UK}
\author{F. Cantalloube}
\affiliation{Max-Planck-Institut f\"ur Astronomie, K\"onigstuhl 17,
     Heidelberg 69\,117, Germany}
\author{W. Brandner}
\affiliation{Max-Planck-Institut f\"ur Astronomie, K\"onigstuhl 17,
     Heidelberg 69\,117, Germany}
\author{Th. Henning}
\affiliation{Max-Planck-Institut f\"ur Astronomie, K\"onigstuhl 17,
     Heidelberg 69\,117, Germany}
\author{C. A. Hummel}
\affiliation{European Southern Observatory, Karl-Schwarzschild-Stra{\ss}e 2, 85\,748 Garching, Germany}
\author{J. Ma\'{\i}z Apell\'aniz}
\affiliation{Centro de Astrobiolog\'{\i}a (CSIC-INTA),
campus ESAC, camino bajo del castillo s/n, 28\,692 Villanueva de la Ca\~nada,
Madrid, Spain}
\author{J.-U. Pott}
\affiliation{Max-Planck-Institut f\"ur Astronomie, K\"onigstuhl 17,
     Heidelberg 69\,117, Germany}
\author{R. Sch\"odel}
\affiliation{Instituto de Astrof\'{\i}sica de Andaluc\'{\i}a (CSIC), Glorieta de la Astronom\'{\i}a S/N, 18\,008 Granada, Spain}
\author{R. van Boekel}
\affiliation{Max-Planck-Institut f\"ur Astronomie, K\"onigstuhl 17,
     Heidelberg 69\,117, Germany}



\begin{abstract}

Characterization of the dynamics of massive star systems and the
astrophysical properties of the interacting components are a
prerequisite for understanding their formation and evolution. Optical interferometry at milliarcsecond resolution is a key
observing technique for resolving high-mass multiple compact systems. Here we report on
VLTI/GRAVITY, Magellan/FIRE, and MPG2.2m/FEROS  observations of the
late-O/early-B type system HD~93\,206~A, which is a member of the
massive cluster Collinder 228 in the Carina nebula complex. With a
total mass of about 90 M$_{\odot}$, it is one of the most compact
massive-quadruple systems known. In addition to measuring the
separation and position angle of the outer binary Aa - Ac, we observe
Br$\gamma$ and HeI variability in phase with the orbital motion of the
two inner binaries. From the differential phases ($\Delta_{\phi}$)
analysis, we conclude that the Br$\gamma$ emission arises from the
interaction regions within the components of the individual binaries,
which is consistent with previous models for the X-ray emission of the
system based on wind-wind interaction. With an average 3-$\sigma$
deviation of $\Delta_{\phi} \sim$15$^{\circ}$, we establish an upper
limit of p$\sim$0.157 mas (0.35 AU) for the size of the Br$\gamma$
line-emitting region. Future interferometric observations with GRAVITY
using the 8m UTs will allow us to constrain the line-emitting regions down to angular sizes of 20 $\mu$as (0.05 AU at the distance of the Carina nebula).

\end{abstract}




\section{Introduction} \label{sec:introduction}

One of the most important observational clues to understand the
formation of massive stars is multiplicity \citep[see
e.g., ][]{Apai_2007, Zinnecker_2007, Peter_2012, Chini_2012}. Some
plausible mechanisms to form massive multiples include: (a) disk fragmentation
\citep{Bonnell_1992, Monin_2007, Krumholz_2009}; (b) the accretion of
wider low-mass systems \citep{Bonnell_2005b,
  Maeder_2002}; (c) formation through failed mergers and stellar
collisions in early dynamical interactions \citep{Zinnecker_2002} and;
(d) disk-assisted capture \citep{Bally_2005}. To discern
which of the aforementioned models should be preferred, an analysis of
the orbital properties in multiple systems (e.g., period distribution, mass ratios, orbital
eccentricities, coplanarity, etc) is required. This is not an easy task, because
companions can cover spatial scales from a few 1/10 Astronomical Units (AU)
to 1000s of AU, and contrast ratios of components can span several orders of
magnitudes. 

\citet{Sana_2014} conducted the first
statistical survey of massive multiple systems combining observations with high
angular resolution facilities (particularly optical interferometry) with archival spectroscopic data. With a main sample of 96 O-stars, this study indicates that $\sim$90\% of the massive stars are
at least binaries and that $\sim$30\% of them belong to a
higher-degree multiple system. For example, the Trapezium system in
Orion \citep{Weigelt_1999, Schertl_2003, Kraus_2009ori} is one of these systems which reveals the existence of stellar companions at multiple spatial scales. In fact, all of its four main components, like
$\theta$1 Ori B \citep{Close_2013},  turn out to be embedded ``micro-clusters'' or
binaries. However, it is not yet clear whether (i) these systems are
stable on long time scales and; (ii) how dynamical interactions
with
the companions can affect their evolution. \citet{Allen_2015}
suggested that quintuplet systems, like $\theta$1
Ori B,  are prone
to dissolve in around $\sim$100 crossing times, with some of the
components merging or ejected \citep[i.e., being the progenitors of
\textit{runaway} stars; see also][]{Fujii_2011}. 

\begin{figure}[thp]
\centering
\includegraphics[width=\columnwidth]{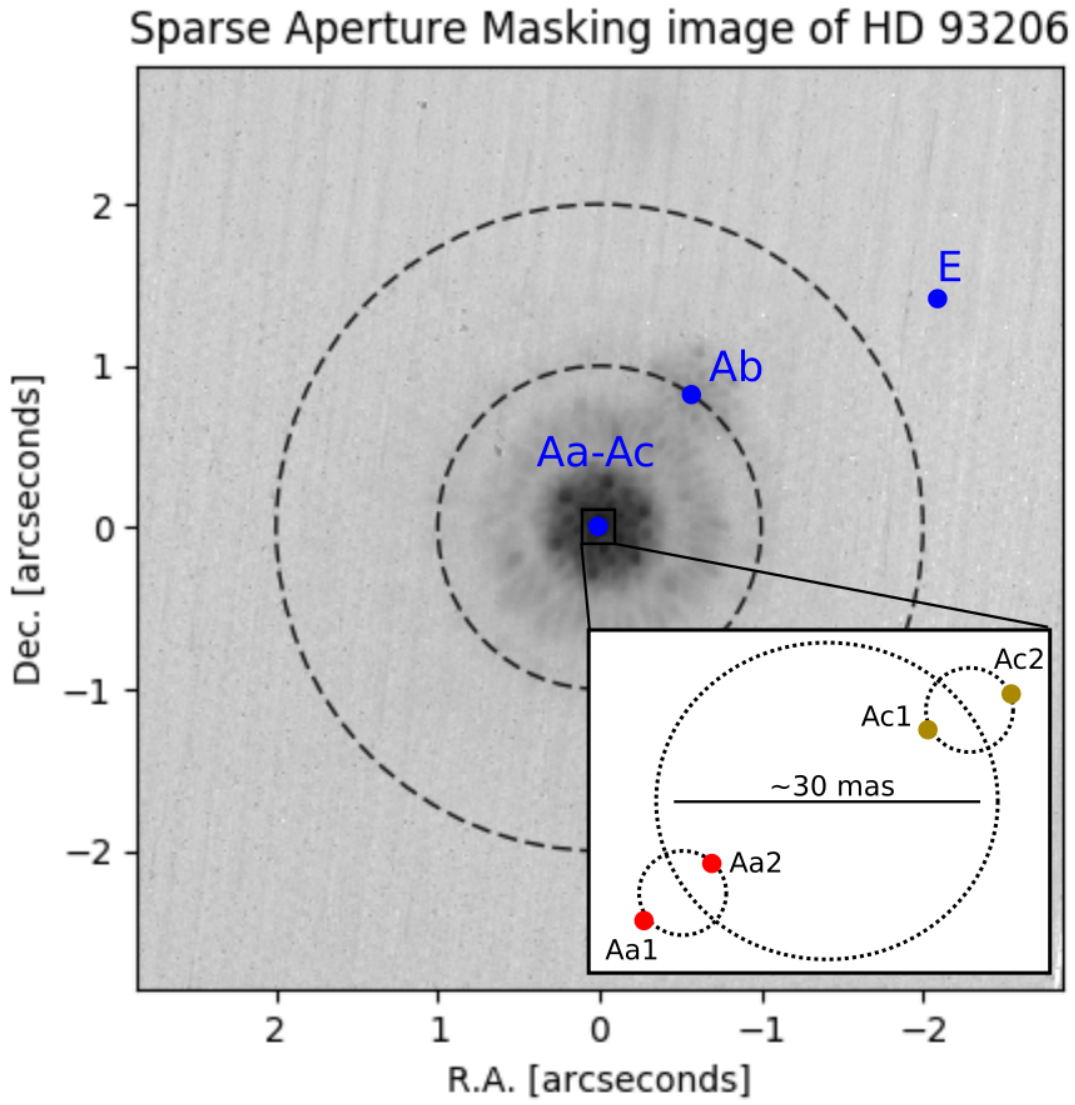}
\caption{ The image displays the Sparse Aperture Masking image of the
  system HD~93\,206. The position of the components Aa-Ac, Ab and E
  are shown with a blue dot. The two concentric
  dashed rings encircle the central 2 and 4 arcseconds around
  HD~93\,206~A. The inset shows the quadruple system
  Aa-Ac, which is the target of this work. The image was taken with
  VLT-NACO in the K$_s$-filter ($\lambda_0$ = 2.2 $\mu$m).}
\label{fig:SAMimg}
\end{figure}

Therefore, resolving
the components of high-degree multiple systems, both in a statistical
way and targeting individual systems,  and monitoring their
orbital motions are mandatory to answer the
aforementioned questions. For this purpose, several high-angular
resolution techniques are required. Among them, infrared long-baseline
interferometry is one of the most crucial techniques to characterize compact systems where the
resolution provided by classical imaging
techniques (e.g., speckle imaging or Adaptive Optics imaging) is
limited. This is particularly true, now that the foremost $K-$band beam combiner
GRAVITY has been successfully deployed at
the Very Large Telescope Interferometer (VLTI). The new capabilities
of GRAVITY, not only offer us the possibility to observe  high-mass
multiples with milliarcsecond resolution, but also at an intermediate spectral
resolution (R$\sim$4000) and higher sensitivity than previous
interferometric instruments. In this study, we present the results of a GRAVITY
spectro-interferometric study of HD~93\,206~A, an X-ray emitter that is known
to be one of the most compact
massive quadruple systems in the Galaxy. The object has been
frequently investigated as a template for the formation of massive multiples.

\begin{table*}
\centering
\caption{Observed stellar pairs in HD~93\,206. Component nomenclature
  follows the WDS (see text). The rest of the information comes from the listed references.}
\begin{tabular}{lccccl}
\hline \hline
Pair  & Year & PA    & sep.      & $\Delta m$ & Reference                \\
     &      & (deg) & (\arcsec) & (mag)      &                          \\
\hline
Aa,Ab & 2014 & 324   & 1.00      &  3.9       & \citet{Sana_2014}       \\
Aa,Ac & 2016 & 328   & 0.03      &  0.4       & This work                \\
A,B   & 2012 & 276   & 7.07      &  5.8       & \citet{Sana_2014}      \\
A,C   & 1934 &  93   & 8.80      &  ---       & \citet{Dawson_1937} \\
A,E   & 2012 & 303   & 2.58      &  7.2       & \citet{Sana_2014}      \\
\hline
\end{tabular}

\label{tab:components}
\end{table*}

\subsection{The source: HD\,93~206~A}

HD~93\,206 ($\equiv$QZ~Car) is a complex multiple system (see Fig.\,\ref{fig:SAMimg}) that is the brightest member of the open cluster Collinder 228, located in the Carina Nebula at a distance of 
2.3 $\pm$ 0.1 kpc \citep{Walborn_1995,
  Smith_2002, Smith_2006}. The known resolved components are listed in
Table~\ref{tab:components}. Before proceeding, some clarifications
are in order:

\begin{itemize}
\item There is some confusion in the literature regarding component
  nomenclature. Here we follow the one of the Washington Visual Double
  Catalog \citep[WDS,][]{Mason_2001}, which is the
      one that deals with the parent-child relationships most
      accurately.
\item  In the current WDS classification, the components of Aa-Ac form a
  quadruple system of 2 spectroscopic binaries with
  Ac being an eclipsing binary. In contrast, the nomenclature of
  \citet{Sana_2014} includes both spectroscopic binaries in component
  Aa, and defines each member with numbers from 1 to 4. 
\item Component E identified in the classification of  \citet{Sana_2014} corresponds to component D in the WDS.
\item Component C could be a spurious detection. It has not been
  observed since 1934, and it is not detected as a 2MASS source. 
\item Components Ab, B, and D (and C if it is real) are significantly
  dimmer ($\Delta_m > $3.0 mag) than Aa and Ac, which dominate the light output of the system and are the only ones bright
      enough to contribute to the integrated spectrum. 

\end{itemize}

\begin{figure}[thp]
\centering
\includegraphics[width=\columnwidth]{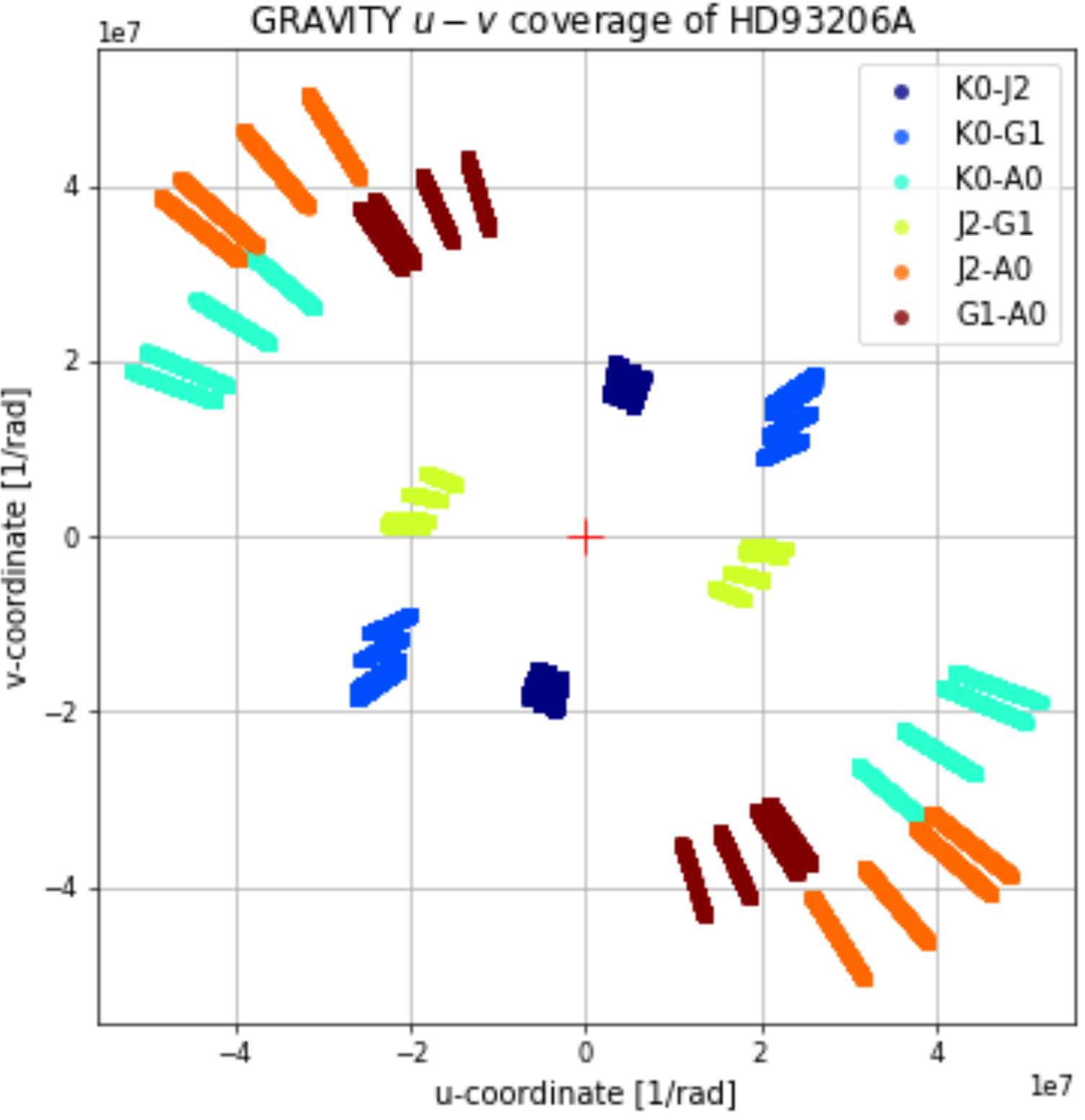}
\caption{ GRAVITY/VLTI $u-v$ coverage of HD~93\,206~A obtained during
  the Science Verification run of the instrument. Four snapshots per baseline were recorded. Baselines are indicated with different colors.The red cross indicates the center of the $u-v$ plane.}
\label{fig:uv_plane}
\end{figure}

This paper deals with the two bright central binaries Aa and Ac (see Fig\,\ref{fig:SAMimg}). Speckle observations with the CHARA speckle camera \citep{Mason_1998} and direct imaging with the Fine Guidance Sensor of the Hubble Space Telescope \citep{Nelan_2004}
could not resolve Aa-Ac, setting an upper limit of $\rho \sim$15.2~mas
($\sim$35~AU) for their separation. However, $H$-band interferometric data taken with PIONIER 
(Precision Integrated-Optics Near-infrared Imaging ExpeRiment) at the Very Large Telescope Interferometer (VLTI) and $K$-band sparse aperture masking (SAM) observations 
obtained with the near-infrared camera NACO at the Very Large
Telescope (VLT) resolved the separation between the two binaries Aa and Ac, finding a projected angular 
distance of $\rho\sim$26 mas (64.4~AU in projection) and with brightness ratios of
$\Delta H = 0.42$ mag and $\Delta K_{\rm S} = 1.18$ mag \citep{Sana_2014}. 

The spectroscopic analysis of the system  \citep{Leung_1979,
  Morrison_1980, Stickland_2000, Mayer_2001} indicates the existence
of two periods of 20.7 d and 6 d for components Aa and Ac, respectively. Therefore, from the spectroscopic point of view, HD~93\,206~Aa-Ac is a quadruple system that can be subdivided into two short-period binary 
systems, Aa1,Aa2 and Ac1,Ac2, in a long-period orbit of $\sim$50~years
around each other. This configuration is different to other hierarchical
quadruple systems \citep[such as HD~17\,505][]{Hillwig_2006, Sota_2011} composed of 
four O-stars in which components are organized in a close pair (with period of days) orbiting a third star with a period of years and a fourth star orbiting around the 
other three with a period possibly measured in millennia. The
intriguing architecture of the system highlights the importance of characterizing the orbital motion of the
components in HD~93\,206~A with the final goal to confront the different
formation scenarios with its dynamical properties. For example, Core
Collapse massive star formation models \citep[e.g.,][]{Krumholz_2009} assume
that companions have coplanar orbital planes, while competitive
accretion models
\citep[e.g.,][]{Bonnell_2006} predicts non-coplanar orbits. 

The spectroscopic analysis also suggests that Ac1,Ac2 is in fact a
semi-detached eclipsing binary, and Aa1,Aa2 is identified as a long-period spectroscopic binary. The individual spectral classification of the stellar components in HD~93\,206~Aa-Ac supports that Aa1 is an O-type supergiant with a spectral type O9.7~Iab or O9.5~I (the classification for the combined spectral type is O9.7~Ibn, \citealt{Sota_2014}). 
Aa2 is undetected in the spectrum but indirect evidence points towards it being several times less massive than Aa1 and with a spectral type in the vicinity of B2~V. 
Ac1 appears to be a giant or supergiant (O8~III or B0~Ib). Although the primary Ac1 is the brightest component, it seems to be less massive than the secondary Ac2 
(possibly O9~V, but its signature on the integrated spectrum is weak),
which suggests a case-B mass transfer within this binary. The whole
system has an estimated 
total mass of $\sim$90~M$_\odot$ \citep{Leung_1979,
  Morrison_1980, Mayer_2001}.

X-ray observations of the quadruple Aa-Ac \citep{Townsley_2011, Broos_2011} revealed an
excess in X-ray emission (average $\sim$7x10$^{-13}$ erg\, s$^{-1}$cm$^{-2}$)
over the expected cumulative bolometric luminosity. A
three component plasma model fitted to the
observed X-ray spectra by \citet{Parkin_2011} supports the
presence of shock heated plasma from wind-wind collision shocks. Those
authors suggest that the system Ac is the main X-ray
emitter. However, the total observed
X-ray flux might be the sum of the emission of the
individual components and the  mutual wind-wind collision
region between the two binaries. If this hypothesis is
correct, we expect to observe infrared shock tracers (like
Br$\gamma$) to be formed in an extended region where the colliding
winds of the two binary systems interact. 

The paper is outlined as follows: Sec.\,\ref{sec:obs}
presents our  GRAVITY/VLTI observations and data
reduction. In Sec.\,\ref{sec:analisis} the analysis of the
interferometric observables and of the source spectrum are
described. In Sec.\,\ref{sec:discussion} our results are discussed and, finally, in Sec.\,\ref{sec:conclusions} the conclusions are presented. 

\section{Observations and data reduction}
\label{sec:obs}

\subsection{GRAVITY observations}

HD~93\,206~A was observed for a total of two hours with GRAVITY
\citep{Eisenhauer_2008, Eisenhauer_2011, Eisenhauer_2017} during
the ESO Science Verification run of the instrument\footnote{ESO
  program: 60.A-9175; PI: J. Sanchez-Bermudez}. Snapshot observations
were obtained at four different hour angles, two of them during June
17th, 2016 and two more during June 18th, 2016. The
observations were carried out using the highest spectral resolution
(R$\sim$4000) of the interferometer in the $K-$band (1.990 $\mu$m --
2.450 $\mu$m), together with the combined polarization and
single-field modes of the instrument. With this configuration, GRAVITY
equally splits the incoming light of the science target between the fringe
tracker and science beam combiners to produce
interference fringes simultaneously in each one of them. While the
science beam combiner disperses the light at the desired spectral
resolution, the fringe tracker beam combiner
always works with a low-spectral resolution of R$\sim$22
\citep{Gillessen_2010} but at a high frequency sampling ($\sim$1 kHz). This allows to correct for the atmospheric piston, 
stabilizing the fringes of the science
beam combiner for up to several tens of seconds.  

\begin{table*}
\caption{GRAVITY/VLTI science target and calibrator data sets.}              
\label{tab:obs}      
\centering                                      
\begin{tabular}{c l c c c c c c c c }          
\hline \hline                       
Date & Target & Type$^{\mathrm{a}}$ & $K_{\mathrm{mag}}$ &
                                                           DIT$^{\mathrm{b}}$
  &  NDIT$^{\mathrm{c}}$ & No. PA$^{\mathrm{d}}$ &
                                                   Airmass$^{\mathrm{e}}$
  & Seeing$^{\mathrm{e}}$ & UD size$^{\mathrm{f}}$ \\    
\hline                                   
    17-06-2016 & HIP~50\,644 & CAL & 4.25 & 30  & 10 & 2 & 1.49, 1.54 &
  0.45, 0.51 & 0.853\\      
        & HD~93\,206~A & SCI & 5.25 & 30  & 10 & 2 & 1.54, 1.67 &
                                                                  0.57,
                                                                  0.54
                          & - \\
      & HD\,94\,776$^{\mathrm{g}}$ & CAL & 3.40 & 10  & 26 & 2 & 1.73,
  1.79 & 0.80, 0.65 & 0.917\\
      & HD~149\,835 & CAL & 4.90 & 30  & 10  & 2 & 1.17, 1.19 & 0.34,
  0.64 & 0.49\\
      & HD~188\,385 & CAL & 5.99 & 30  & 10 & 2 & 1.45, 1.51 & 0.47,
                                                               0.43 & 0.224 \\
\hline
    18-06-2016 & HD~93\,206~A & SCI & 5.25 & 30  & 10 & 2 & 1.40, 1.43
  & 1.0, 0.77 & - \\      
       & HD~94\,776$^{\mathrm{g}}$ & CAL & 3.40 & 10  & 26  & 2 &
                                                                   1.45,
                                                                   1.47
  & 0.57, 0.53 & 0.917 \\
       & HD~166\,521 & CAL & 4.52 & 30  & 10 & 2 & 1.06, 1.07 & 0.48,
  0.48 & 0.578\\
       & HD~164\,031 & CAL & 4.11 & 30  & 10 & 2 & 1.31, 1.37 & 0.59,
  0.88 & 0.707\\
       & HD~8315 & CAL & 4.04 & 30  & 10 & 1 & 1.18 & 0.78 & 0.749\\

\hline                                             
\end{tabular}
\begin{list} {}{} \itemsep1pt \parskip0pt \parsep0pt \footnotesize
\item[$^{\mathrm{a}}$] This parameter indicates the purpose of the target observed. CAL means that the source corresponds to a calibrator star and SCI means that the source corresponds to our science target.
\item[$^{\mathrm{b}}$] Detector integration time in seconds.
\item[$^{\mathrm{c}}$] Number of frames to estimate the complex
  visibilities per data cube.
\item[$^{\mathrm{d}}$] Number of observed position angles per night.
\item[$^{\mathrm{e}}$] Airmass and Seeing per target data set.
\item[$^{\mathrm{f}}$] $K-$band estimated Uniform Disk size in
  milliarcseconds obtained from \texttt{SearchCal}
  \citep{Bonneau_2006, Bonneau_2011}.
\item[$^{\mathrm{g}}$] K0III star used to calibrate the HD~93\,206~A spectrum.
\end{list}
\label{tab:obs}
\end{table*}

All the HD~93\,206~A data were recorded with the
A0-G1-J2-K0 configuration of the 1.8-meter Auxiliary Telescopes (ATs) of
the VLTI. This configuration provides a maximum projected baseline of
$\sim$120 meters (J2-A0) and a minimum one of $\sim$39 meters (K0-J2)
for the target coordinates. These baseline lengths correspond to a
maximum angular resolution ($\theta$=$\lambda$/2B) of $\theta
\sim$1.89 mas and a minimum one of $\theta \sim$5.8 mas, at a
central wavelength of $\lambda_0$= 2.2 $\mu$m. Figure
\ref{fig:uv_plane} displays the science target $u-v$ coverage obtained
with the aforementioned GRAVITY observations. 

The interferometric observables (squared visibilities, closure phases,
and differential phases) as well as the source's spectrum
were obtained using the version 0.8.4 of the instrument's data reduction
software\footnote{http://www.eso.org/sci/software/pipelines/gravity/gravity-pipe-recipes.html}
provided by ESO \citep{Lapeyrere_2014}; it uses of a series of
\texttt{esorex}\footnote{http://www.eso.org/sci/software/cpl/esorex.html}
recipes to estimate the raw observables and calibrate them.  The data
reduction process converts the science camera frames \citep[where the interference fringes are recorded using the ABCD sampling
method, ][]{Colavita_1999} into a set of 1D-spectra from
which the observables are estimated. 

First, the science detector
is characterized by measuring its bias, pixel gain, bad
pixel and wavelength maps. The wavelength maps of the Fringe
  Tracker and Science beam combiners are calibrated using the GRAVITY
  Calibration Unit \citep{Blind_2014}. For the Science Beam Combiner,
  the fiber wavelength scale is obtained through observations of the
  internal Halogen lamp having the Fiber Differential Delay Lines in
  close-loop and fringe tracking. At the same time, the calibration
  unit delay line positions are modulated, while their position is
  monitored using the internal laser diode ($\lambda_0$ = 1.908
  $\mu$m). The measured fringe phase shifts are converted to
  wavelength knowing the introduced OPD offsets, with a wavelength
  precision of $\Delta \lambda$ = 2nm. An Argon spectral calibration lamp (which provides 10 lines in the GRAVITY wavelength range) is used to derive the vacuum wavelength scale and correct the wavelength map of the fiber dispersion, achieving an absolute wavelength calibration equivalent to 1/2 of spectral element at the highest spectral resolution of the instrument (0.1 nm, equivalent to a relative uncertainty of $5 \times 10^{-5}$ in high spectral resolution mode).  

The transfer function of the science beam combiner is
calibrated by using the P2VM algorithm \citep{AMBER_Tatulli_2007,
  Lacour_2008}. It, first, requires to determine the V2PM matrix, as
defined in \citet{Lapeyrere_2014}, for each spectral bin. The internal Halogen lights are used to
calibrate this matrix in three steps: first, for the power
transmission of each one of the telescopes; second, by computing the visibility amplitude and phase for each baseline (i.e.,
internal contrast and phase) produced by an induced optical path delay
(OPD) modulation and; third, by computing the closure phases. Finally, the inverse of the
V2PM matrix is used to extract the interferometric observables from
on-sky observations. The P2VM calibration files are
generated from the raw data with the \texttt{esorex} recipe \texttt{gravity\_p2vm}. Once they are
created, the \texttt{gravity\_vis} recipe can be used to extract the
total flux per telescope and correlated flux per baseline (i.e., the
raw observables). The recipe \texttt{gravity\_vis} includes a bootstrapping method to
compute the complex visibility and bispectrum errors, as well as a
visibility-loss correction factor due to the integration time difference between the
fringe tracker frames and the science ones. 

To calibrate the interferometric observables, interleaved observations
of the science target and point-like sources were
performed. Table \ref{tab:obs} displays the observational setup used
for the science target together with the different calibrators. The interferometric calibration was
performed using the \texttt{gravity\_viscal} recipe. This routine
computes the instrumental transfer function by correcting the observed
calibrator visibilities by the theoretical ones according to the
estimated angular size of the calibrator. The algorithm then groups calibrators with the
same observational setup as the science target and interpolates their 
transfer functions to the acquisition time of the science target observations. Finally,
the recipe corrects the target raw visibilities for the estimated
transfer function, delivering the calibrated observables. 

Before analyzing the data,
those V$^2$ points with a SNR $\leq$ 5 and
closure phases with $\sigma_{cp} \geq$ 40$^{\circ}$ were
flagged. Figures \ref{fig:snr_v2} and
\ref{fig:snr_cps} in the Appendix \ref{sec:SNR_obs} display the
distributions of the V$^2$ SNR and of the errors in the closure
phases after flagging. From these distributions, it is observed that
three baselines have a considerably small SNR, particularly in the second data set
where only a few points remain after flagging. These baselines are
connected with the K0 station, which shows a poor transmission throughput at the time of the Science Verification observations. The first data set has the
highest quality mainly because of the excellent weather conditions
under which it was taken (see Table \ref{tab:obs}). 

\begin{figure*}
\centering
\includegraphics[width=\linewidth]{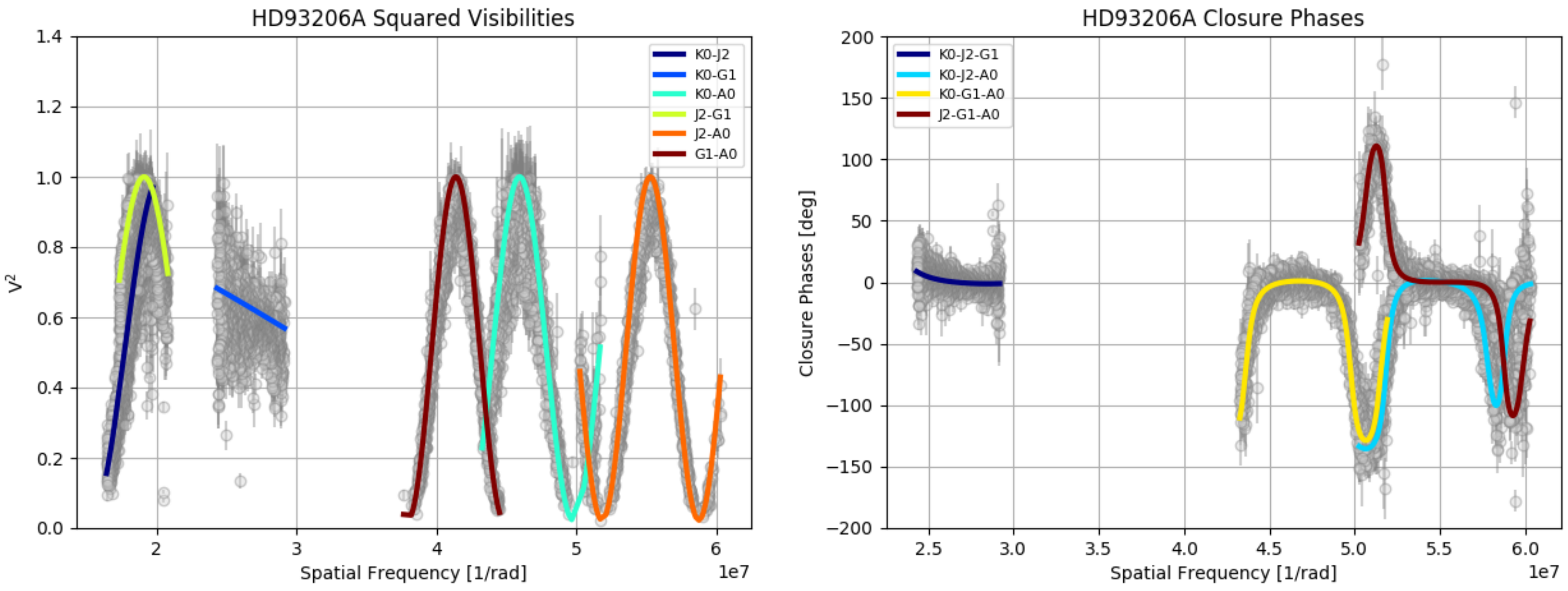}
\caption{ Calibrated V$^2$ visibilities and closure phases versus
  spatial frequency. One data set of the GRAVITY/VLTI observations
  is displayed with gray dots (MJD: 57557.046). The best-fit average model
  is overplotted with color lines. Baselines are indicated with different
  colors and their corresponding telescope stations displayed on the
  frames.}
\label{fig:model_vis}
\end{figure*}

\begin{figure}
\centering
\includegraphics[width=\columnwidth]{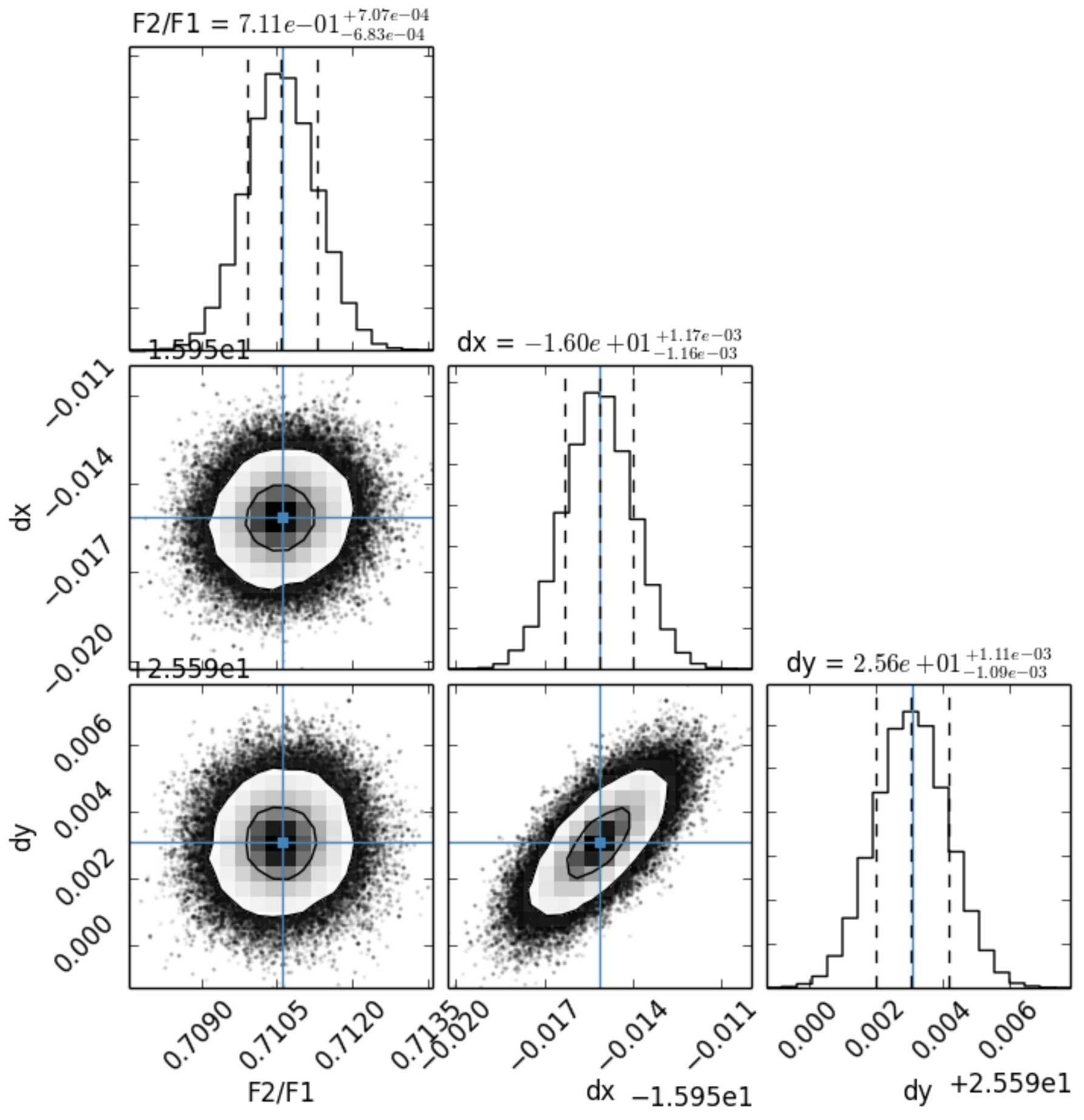}
\caption{Posterior distributions of the fitted parameters. The 2D
  distributions shows 1 and 2 standard deviations encircled by a
  black contour. The position in the distributions of the best-fit solution found with the non-linear least
squares method is displayed with a blue square. The 1D histograms show
the expected value, $\mu$, and $\pm$1$\sigma$ with vertical dashed
lines, together with their corresponding values at the top. }
\label{fig:triangle}
\end{figure}

\subsection{FIRE observations }
\label{sec:fire_obs}

In order to determine the K-band spectral features present in the HD~93\,206~A
spectrum,
we have obtained near-infrared spectra using the
Folded-port InfraRed Echellette \citep[FIRE; ][]{Simcoe_2013} spectrograph,
attached to the 6.5 m Baade Magellan Telescope at Las Campanas Observatory.
Observations were gathered on March 28th, 2016 (HJD 2457475.66).
We have obtained four one second exposure at airmass 1.17,
under excellent seeing conditions (0.5 arcsec).
FIRE was used in echelle mode, and the selected
slit width was 0.6 arcsec. The FIRE spectrum is spread over 21 orders
from 0.82 to 2.51 microns, providing a resolving power R$\sim$6000. The star observations were followed by a bright telluric standard
of spectral type A0V, in this case HD~83\,280.

Data reduction and the spectral extraction were performed with the FIREHOSE
package. It is based on the MASE \citep{Bochanski_2009} and SpeXtool
packages \citep{Vacca_2003, Cushing_2004}.
At the beginning of the  night, we have also obtained dome flats and
Xenon flash lamps
to build a pixel-to-pixel response calibration. 
For wavelength calibration we used Thorium-Argon lamp exposures
taken immediately after the target exposures, and OH sky
emission lines extracted from the target exposures.
The telluric correction was accomplished
using the methods and routines developed in SpeXtools.

Adopting the ephemeris published by \citet{Mayer_2001} for Aa-Ac,
the FIRE spectrum of  HD~93\,206~A corresponds to phase 0.97 of the orbit of Ac,
that is in the primary eclipse of the system Ac, and phase 0.74,
which corresponds to the maximum radial velocity for the primary (Aa1)
in the pair Aa. We should take into account that the errors reported in the
ephemeris of both systems could lead to phase errors of about 3\% for system Ac
and 15\% for system Aa. 

\subsection{FEROS observations}

Optical spectroscopic results are presented in
Section\,\ref{sec:discussion}. The data were obtained as part of the OWN Survey \citep{Barba_2010}. They were gathered using the FEROS
spectrograph attached to the 2.2 m MPG/ESO telescope at La Silla, in May
2-5th, 2009 and May 21th, 2012. These FEROS observations were reduced
with the pipeline provided by ESO in the MIDAS environment.

\section{Data analysis}
\label{sec:analisis}

\subsection{Modeling the interferometric observables }
\label{subsec:model_fitting}

\subsubsection{MCMC modeling }
\label{subsec:mcmc_fitting}

In the calibrated V$^2$ and closure phases, the target presents the prototypical cosine signature
of a resolved binary system. Figure \ref{fig:model_vis}
displays one of the GRAVITY data sets (MJD:57557.046) where the binary
signature is clearly visible. As expected, with the current GRAVITY
observations we could only map the separation between the two
components Aa and Ac, but not resolve the individual components of
these binaries (which have angular separations considerably less than 1 mas). A
geometrical model fitting of a binary composed of two point-like
sources was applied to the
observables. The expression used to estimate the complex visibilities,
$V\mathrm{(u,v)}$, is the following one:

\begin{equation}
V\mathrm{(u,v)} = \frac{1+\mathrm{F_{Ac}/F_{Aa}}G(\zeta)e^{-2\pi j
    \mathrm{(\Delta\theta_x u + \Delta\theta_y v)}}}{1 +
  \mathrm{F_{Ac}/F_{Aa}}}\,,
\end{equation}

with, 

\begin{equation}
G(\zeta)=|\frac{sin\zeta}{\zeta}|\,\, \,\, \,\,\mathrm{with}\,\, \,\, \,\,\zeta=\frac{\pi\mathrm{(\Delta\theta_x u + \Delta\theta_y v)}}{R}\,,
\end{equation}

where $\mathrm{F_{Ac}/F_{Aa}}$ is the flux ratio between the components of
the binary, (u, v) are the components of the spatial
frequency sampled per each observed visibility point
(u=B$_x$/$\lambda$ and v=B$_y$/$\lambda$), and ($\Delta\theta_x$, $\Delta\theta_y$) are
the East-West and North-South components of the projected angular separation
between primary and secondary components of the binary. $G(\zeta)$ is
a correction factor due to the bandwidth smearing of the finite
bandpass used and R=$\lambda/\Delta \lambda$ is the spectral
resolution of the interferometer (in this case R$\sim$4000). The total
flux of the binary is normalized to unity such that F$_{Aa}$+F$_{Ac}$=1.

V$^2$ and
closure phases across the entire bandpass were simultaneously used to
estimate the average $\mathrm{F_{Ac}/F_{Aa}}$ and  ($\Delta\theta_x$,
$\Delta\theta_y$)  of the
binary. The fitting
process was performed using a dedicated Markov chain Monte-Carlo (MCMC) routine based on
the python software \texttt{emcee} \citep{Foreman_2013}. First, an approximation of the best-fit parameters was
obtained using the non-linear least squares minimization algorithm
inside the python library
\texttt{scipy.optimize}\footnote{http://www.scipy.org/}.  To account
for the standard deviations and correlations of the model parameters,
we explore the solution space around the best-fit values obtained with
the least squares method; 300 random points with a linear distribution between
$\pm$10$\sigma$ were generated for each one of the parameters. For every random point, 700 iterations were
performed using the MCMC algorithm to maximize the posterior probability
of the model. The $\chi^2$ of the best-fit model is $\chi^2=$ 3.93, while the parameter
values and three-sigma uncertainties of the best-fit model are reported
in Table \ref{tab:model_vis}. Figure \ref{fig:triangle} shows the posterior probability distributions of
the parameters, their correlations, as well as their individual marginalized
distributions in 1D histograms along the diagonal of the plot.

\begin{table*}
\caption{Best-fit geometrical model with the different algorithms used
in this work.}              
\centering                                      
\begin{tabular}{l c c c c c c c c}          
\hline \hline                     
 &  \multicolumn{2}{c}{MCMC fit} & \multicolumn{2}{c}{\texttt{LitPro}
                                   fit} & \multicolumn{2}{c}{\texttt{CANDID}
                                   fit}  & \multicolumn{2}{c}{Average
                                   fit} \\
Parameter & Value & $\pm 3\sigma$ & Value & $\pm 3\sigma$ &
                                                                      Value
                  & $\pm 3\sigma$ & Value & $\pm 3\sigma$ \\      
\hline                             
    $\mathrm{F_{Ac}/F_{Aa}}$ & 0.710 & $\pm$0.002 & 0.67 & $\pm$0.11 & 0.832 &
                                                                    $\pm$0.004
  & 0.740 &  $\pm$0.14 \\ [0.1cm]     
      $\Delta \theta_x$ [mas] & -15.965 & $\pm$0.0013 & -15.91 & $\pm$0.020 &
                                                                   -15.942
                                &  $\pm$0.0054 & -15.940 & $\pm$0.045 \\[0.1cm]
     $\Delta \theta_y$ [mas] & 25.593 & $\pm$0.0034 & 25.645 & $\pm$0.024 &
                                                                 25.591
                                & $\pm$0.005 & 25.609 &
                                                                    $\pm$0.054 \\[0.1cm]
\hline                                             
\end{tabular}
\label{tab:model_vis}
\end{table*}

\begin{figure}
\centering
\includegraphics[width=\columnwidth]{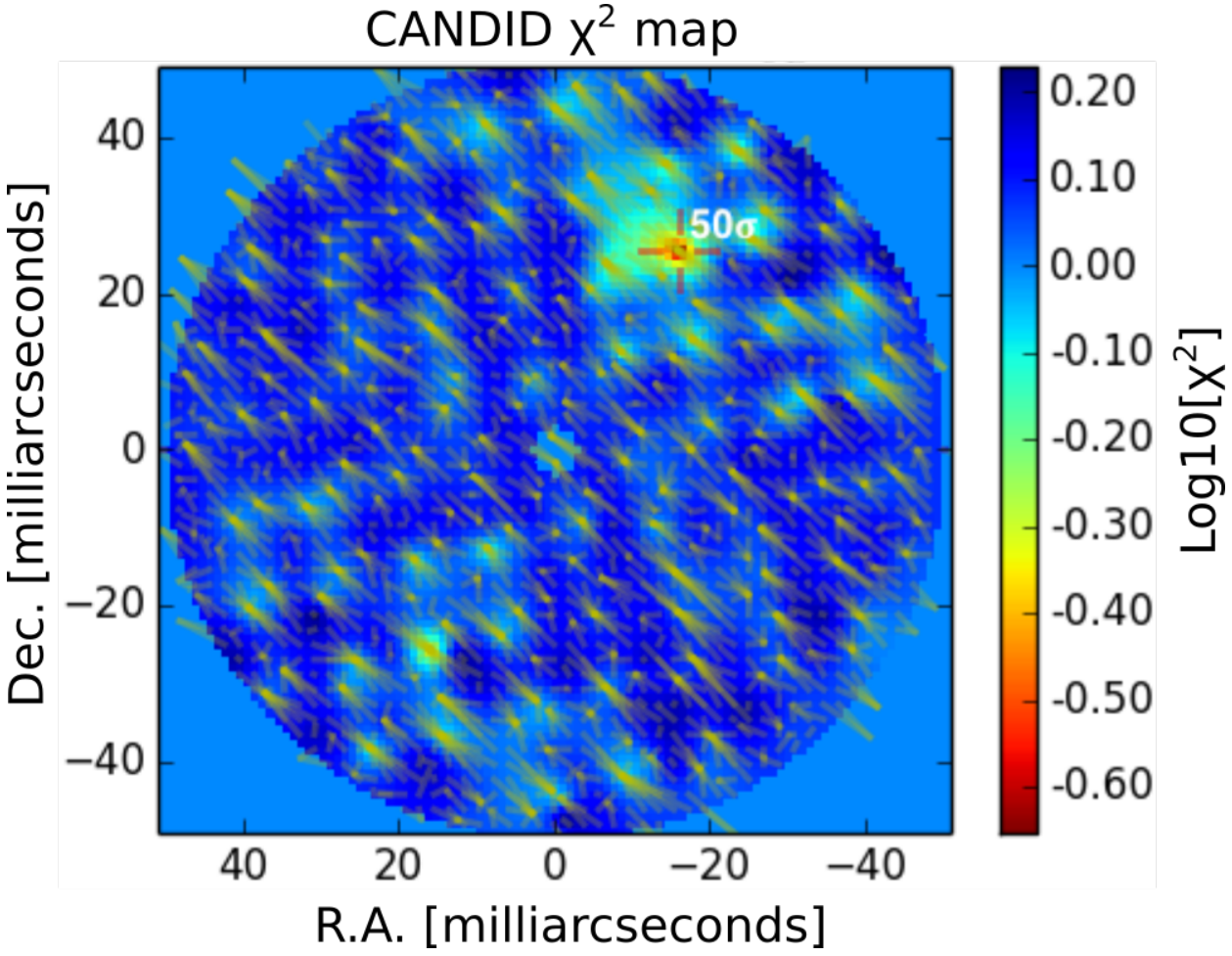}
\caption{The figure displays the \texttt{CANDID} interpolated map of the $\chi^2$ in a
field-of-view of 100 mas. The position of the secondary is marked with a red
cross. The color scale is logarithmic. The yellow lines indicate the
convergence vectors (from the starting points to the final fitted
position) of the Levenberg-Marquardt algorithm for each one of the points sampled in the grid.  }
\label{fig:CANDID_fitmap}
\end{figure}

\subsubsection{\texttt{LitPro} modeling}

Additionally, to our MCMC modeling, we also performed the analysis of the
binary using \texttt{LitPro}\footnote{Available at: http://www.jmmc.fr/litpro} \citep{LITpro}. This is a code that uses a
gradient descent method to perform the optimization; it is based on a
Levenberg-Marquardt algorithm combined with a Trust Region method. The software allowed us to use
several pre-defined geometrical components to represent the interferometric
data. In this case, we use a similar setup to the MCMC model (a binary
composed of two point-like sources). There are two main differences between the
two methods: (i) the first one is the lack of a correction in \texttt{LitPro} due to
bandwidth smearing; (ii) the second one is the flagging method applied
by \texttt{LitPro}, which masks $V^2 <$ 3$\sigma_{V^2}$ or $V^2 >$ 1+ 3$\sigma_{V^2}$ . Table\,\ref{tab:model_vis} displays the best-fit
model obtained with this software with an achieved
$\chi^2$=3.89. Despite the differences in the algorithms and their
setup, the parameters obtained with MCMC and \texttt{LitPro} are in
very good agreement.

\begin{figure*}
\centering
\includegraphics[width=17 cm, height=7 cm]{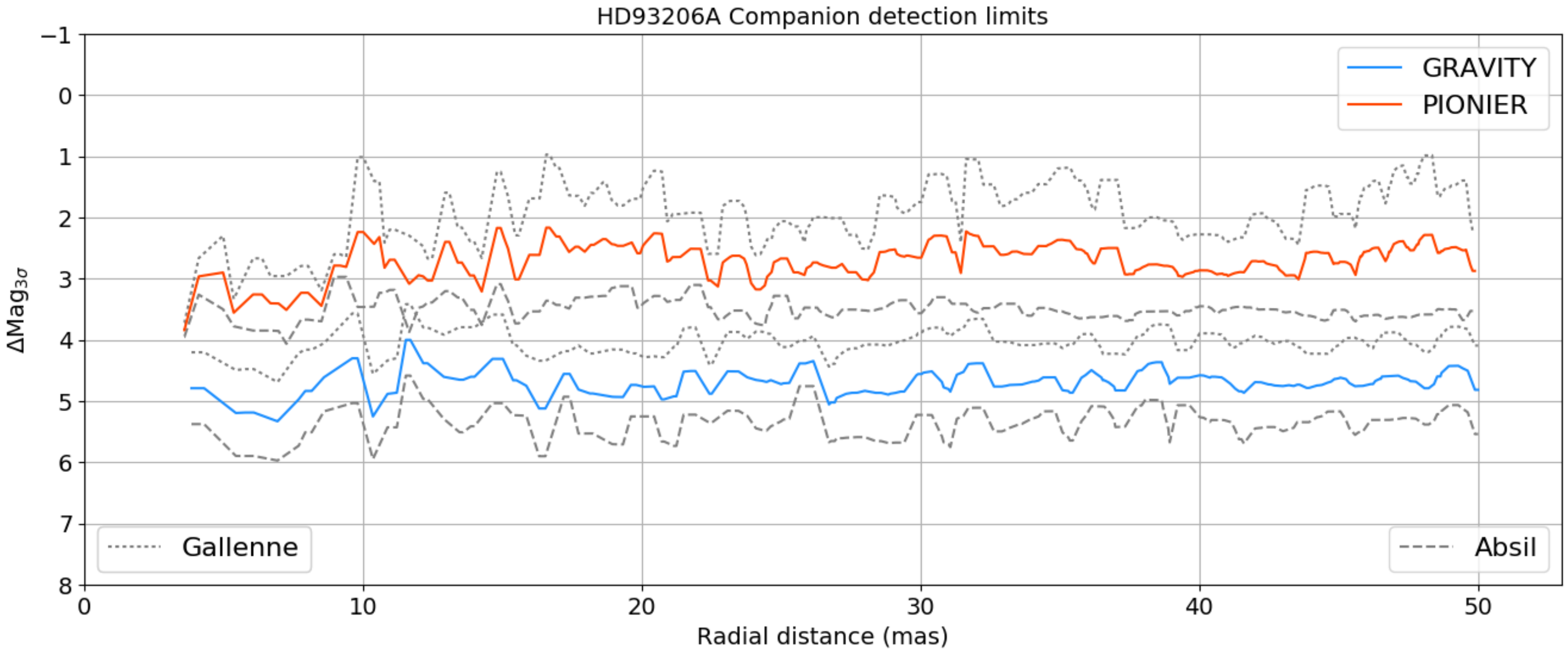}
\caption{3$\sigma$ contrast limits of secondary components are plotted for separations between 3 and 50 mas in the GRAVITY and PIONIER data of HD~93\,206~A. The red and blue solid lines indicate the
  average detection limits between the two detection methods used. The
response of each one of the detection algorithms is plotted in dotted
and dashed gray lines for the \citet{Absil_2011} and
\citet{Gallenne_2015} methods, respectively. }
\label{fig:CANDID_detection}
\end{figure*}

\subsubsection{\texttt{CANDID} modeling and detection limits}

We also
used \texttt{CANDID}\footnote{Available at: https://github.com/amerand/CANDID} \citep{Gallenne_2015} to analyze the
interferometric data of HD~93\,206~A. This Python code uses an
Levenberg-Marquardt algorithm to fit for a binary system performing a
systematic exploration over a given grid. The model of the binary
represents a spatially resolved primary component (to which a uniform
disk is fitted), as well as the bandwidth smearing of the visibilities
\citep[see ][]{Lachaume_2013}.  In this case, we considered the two
components of the binary to be unresolved and the size of the binary
component was set to zero. For the grid search, we
selected a step of p=1.89 mas, which is equivalent to $\lambda$/2B$_{max}$. Additionally to the binary
fitting, \texttt{CANDID} also allows to obtain an
estimation of the 3$\sigma$ detection limit for companions at different
separations from the primary. The code includes the algorithms
described in \citet{Absil_2011} and \citet{Gallenne_2015} to look for
high-contrast companions in the searched grid. 

Table\,\ref{tab:model_vis} displays the best-fit parameters obtained with
\texttt{CANDID}. Figure\,\ref{fig:CANDID_fitmap} displays the map of the $\chi^2$
minima of the binary model with the position of the secondary marked with a red cross. In
the case of our data, the secondary was detected with the highest
confidence value allowed by \texttt{CANDID} (50$\sigma$). The minimum $\chi^2$ achieved by the
searching algorithm is $\chi^2$ =2.46. The flux ratio and position of the secondary 
found with \texttt{CANDID} agrees with
the values obtained with the other two methods previously described
(see Table\,\ref{tab:model_vis}). 

\texttt{CANDID} was also used to estimate the maximum
3$\sigma$ magnitude difference, $\Delta$Mag$_{3\sigma}$, to detect a companion for a separation
range between 3 and 50 mas using both detection methods available in
the code. Here, we only used one of the GRAVITY data sets (MJD:57557.046), to properly
compare the results with the 2012 PIONIER data set reported by \citet{Sana_2014}. The published PIONIER astrometric position and flux
ratio were adopted to subtract the secondary component from the data
and search for faint companions in the grid with a step of p=1.4 mas.
Figure\,\ref{fig:CANDID_detection} displays the resulting detection limits. The blue and red solid lines indicate the average
detection limits between the two detection methods available in
\texttt{CANDID}. These results support that the maximum detectable magnitude
difference in the PIONIER data is $\sim$3 mag, while for GRAVITY it
goes up to $\sim$5 mag. 

As evident in Table\,\ref{tab:model_vis}, the model parameters
obtained with the three different methods agree
quite well. However, the standard deviation obtained for
each of the algorithms appears to be underestimated. This is
mainly caused by slight
deviations in the used expressions for the optimization, as well as
from the flagging methods applied in each one of the methods. To obtain a
more robust estimation of the best-fit parameters and their uncertainties, we computed the
mean and the standard deviation among the best-fit models determined
with the different optimization methods. These estimates are also
reported in columns 8 and 9 in Table\,\ref{tab:model_vis}, and those are the values
adopted for the following analysis in this work. Figure\,\ref{fig:model_vis}
displays (for clarity) one of the GRAVITY data sets together with the
best-fit average model overplotted. Notice how the trend in
V$^2$ and CPs are well reproduced by the best-fit model.

\subsection{Extracting the source spectrum}

Together with the interferometric observables, GRAVITY delivers the
spectrum of the source for each of the four telescopes that forms the
interferometer. The \texttt{esorex} data reduction software delivers
the spectrum flattened by the instrumental transfer
function. To correct for atmospheric effects, it is
necessary to use a spectroscopic calibrator. Here, the K0III star
HD~94\,776 was used (see Table \ref{tab:obs}). The spectral type of
this star does not exhibit strong absorption or emission lines between
2.0 and 2.29 $\mu$m. However, it shows a CO band head in absorption
from 2.29 $\mu$m on. Therefore, wavelengths larger than 2.29 $\mu$m
were discarded from the spectral analysis. 

The calibrated spectrum of HD~93\,206~A was obtained by dividing
each one of the four raw-science spectra per data set (telescope) by its
corresponding raw-calibrator spectrum. Every science data set
was corrected by each one of the calibrator observations per
night, obtaining a total of 32 samples of the source spectrum. After the ratio was computed, the different calibrated spectra
were normalized with a 3rd degree
polynomial fitted to the continuum. Due to the presence of outliers in
the different spectra (associated to bad pixels) and to the
difference in the SNR caused by different telescopes transmissions, a Principal
Component Analysis (PCA) was used to obtain a median spectrum of the source.  
 
The PCA algorithm \citep[see][]{Jolliffe_1986} filters out the high-frequencies in the
data that correspond mostly to noise, allowing us to keep the components of
the signal that maximize the variance (i.e., the direction with the
maximum signal) between the different samples of
the HD~93\,206~A spectrum. The PCA algorithm \citep{Astro_statistics_2014} consists of the following
steps:

\begin{itemize}
\item We center each one of the spectra in the set of data $\{x_i\}$ by
  subtracting the mean spectrum.
\item A matrix $\mathbf{X}$ of $N \times K$ dimensions was built with
  the previously subtracted spectra $\{x_i^{\mathrm{sub}}\}$. Here, $N$ corresponds to the number of samples of the
  HD~93\,206~A spectrum (i.e., $N$=32) and $K$ to the number of dimensions in the
  spectrum (i.e., to the spectral bins in the data). 
\item PCA aims at projecting $\mathbf{X}$ into a space
  $\mathbf{Y}=\mathbf{X}\mathbf{R}$, where $\mathbf{R}$ is aligned to
  the direction of maximal variance. This projection is done through
  a single value decomposition (SVD) of $\mathbf{X}$ itself in its
  eigenvalues and eigenvectors.  
\item The reconstruction of each individual spectrum, ${x_i}$, is given by:

\begin{equation}
\mathbf{x}_i(k)=\mathbf{\mu} (k) + \sum_j^{r < R} \theta_{ij}\mathbf{e}_j(k)\,,
\end{equation}

where the indices $k$ corresponds to the different wavelength bins, $i$
to the number of the input spectrum and $j$ to the number of the
eigenvector used. $\mathbf{\mu}$ corresponds to the mean of the original spectra per
wavelength bin. R is the total number of eigenvectors $\mathbf{e}(k)$. In this case,
the summation was performed only up to the coefficient $r$ which
corresponds to a cumulative variance of 50\% in the data. The
coefficients, $\theta_{ij}$, are given by:

\begin{equation}
\theta_{ij}= \sum_k \mathbf{e}_j(k)\mathbf{x}_i^{\mathrm{sub}}\,,
\end{equation}
\end{itemize} 

\begin{figure*}
\centering
\includegraphics[width=\linewidth]{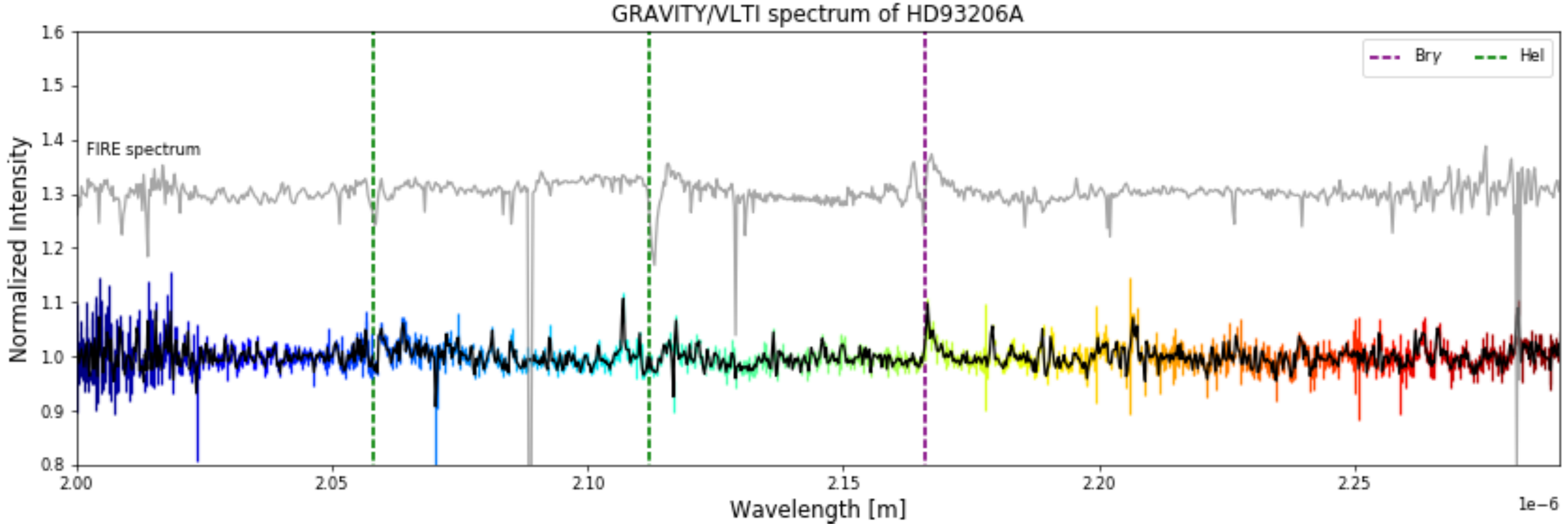}
\caption{The figure displays the HD~93\,206~A normalized spectrum obtained with GRAVITY. Different spectral bins are shown with
  different colors. Some of the main spectral features are identified
  and plotted with different colored dashed lines (see label on the
  plot). The FIRE spectrum is overplotted in gray and has been shifted from unity for better comparison
  with the GRAVITY one.}
\label{fig:spectrum}
\end{figure*}

Figure \ref{fig:spectrum} displays a normalized median of the
HD~93\,206 A spectrum after the
PCA analysis. The displayed error bars correspond to the standard
deviation of the median. Although no prominent spectral features are
observed, some of them, like Br$\gamma$ and HeI, are identified. For comparison, a normalized spectrum of the source taken
with the FIRE spectrograph, at the 6-meter Magellan telescope, is also
plotted. Similar spectral features are distinguished in both spectra.

\section{Discussion }
\label{sec:discussion}

From the geometrical model, a separation between components in the
outer binary of $\rho$=30.16$\pm$0.02 mas and a position angle of $\theta$=328.09$\pm$0.03$^{\circ}$ were
derived. These estimates agree with the values reported by
\citet{Sana_2014} using PIONIER ($\rho$=25.76$\pm$0.54 mas;
$\theta$=331.31$\pm$1.45$^{\circ}$) and SAM ($\rho$=28.05$\pm$5.41 mas; $\theta$=331.02$\pm$9.61$^{\circ}$) observations
taken in 2012, but with a precision between one and three orders of magnitude
better, respectively. The SAM data derived a $\Delta K_s$=1.18$\pm$0.20 mag,
however, our GRAVITY observations suggest a $\Delta K$=0.33$\pm$0.02 mag; a value
that is closer to the  $\Delta H$=0.42$\pm0.18$ mag
derived with the PIONIER observations. The angular separation between
components of the binary is at the limit of the SAM angular-resolution
($\theta_K \sim$35 mas; for 8-meter telescopes). Thus, this could explain the poor performance of such
technique to reliably recover the $\Delta K$ and separation of the
system Aa-Ac. In addition, it highlights the importance of long-baseline
interferometric observations, like our GRAVITY data to
cover the relevant angular scales of systems like HD~93\,206~A. 

\begin{figure}
\centering
\includegraphics[width=8 cm]{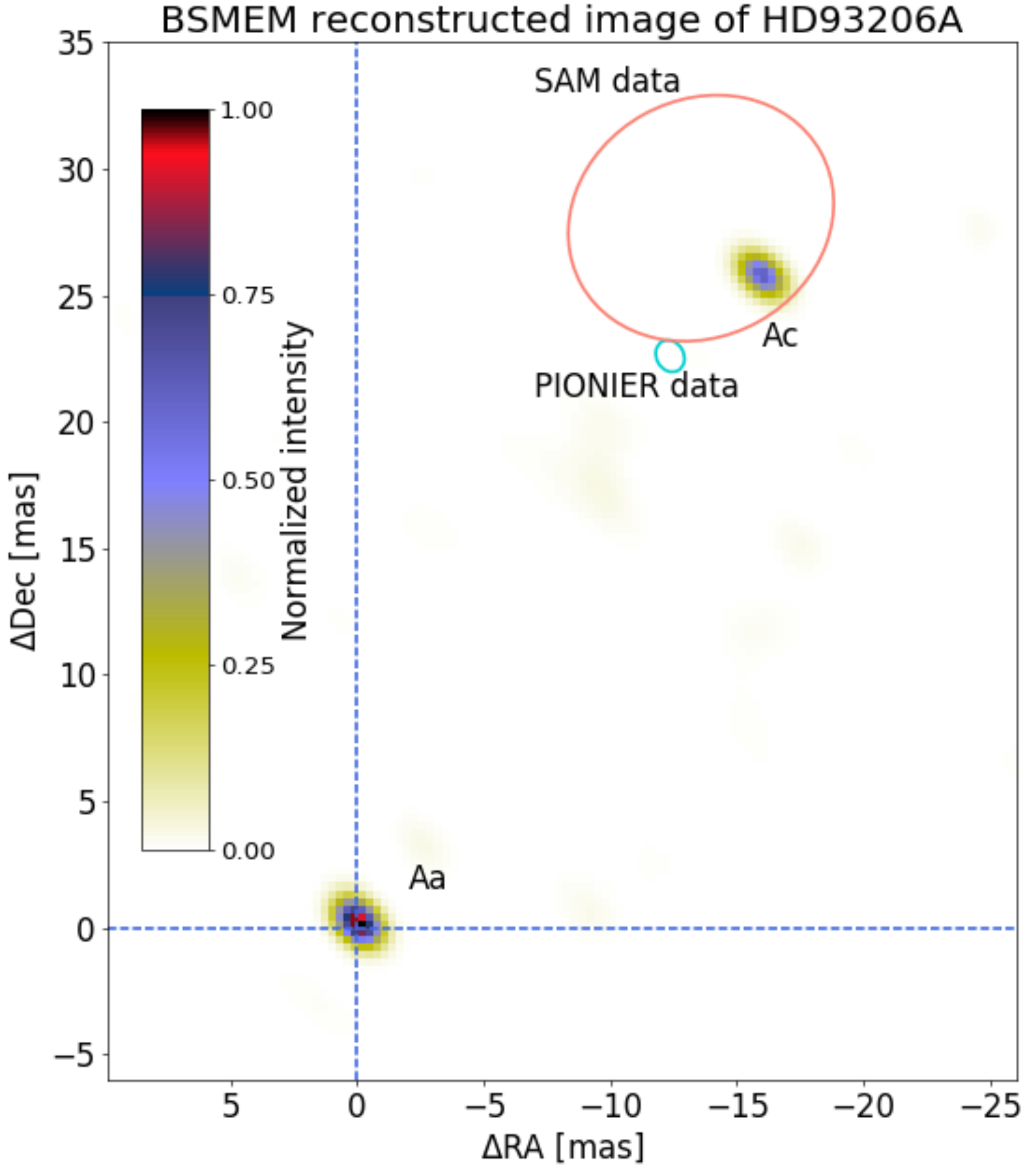}
\caption{Reconstructed BSMEM image of HD~93\,206~A. The components of
  the outer binary Aa-Ac are labeled on the figure. The color
  scale is also shown. Cyan and a red ellipses show the
  positions (within $\pm$1$\sigma$) of the Ac component according to the
  PIONIER and NACO/SAM values reported by \citet{Sana_2014},
  respectively. Notice the large uncertainty ellipse of the SAM data
  compared to the precision of the long baseline interferometric
  data. The positional difference between the PIONIER and GRAVITY epochs is due to orbital motion.}
\label{fig:bsmem_im}
\end{figure}

With the position angles derived in 2012 and in this work, a projected
North-West motion in the plane of the sky of component Ac relative
to Aa is detected. Figure \ref{fig:bsmem_im}
displays a BSMEM \citep{Buscher_1994, Baron_2008} reconstructed image
from our GRAVITY
data together with the 2012 PIONIER and SAM positions marked on it. With only two
astrometric points, it is not possible to derive a reliable orbital
solution of the outer binary Aa-Ac (see
Fig.\,\ref{fig:bsmem_im}). Nevertheless, (i) assuming a
total mass of 90 M$_{\odot}$ \citep{Leung_1979, Morrison_1980}, (ii) a projected
circular orbit with a mean separation at the aphelion of $\sim$30
mas, (iii) and a parallax to the
Carina Nebula $\epsilon$ = 0.43 $\pm$ 0.02 \citep{Smith_2002}, a rough
orbital period of P = 61 $\pm$ 4 a is inferred. However, this
estimate highly depends on the orbital elements of the system (e.g.,
eccentricity, inclination, mass, etc). For example, taken the previous constraints into account, a system projected in the plane of the sky with
eccentric orbits e=0.5 and e=0.9, will result on P= 33 $\pm$ 2 a and P = 23 $\pm$ 1.6 a,
respectively. These, still, poor constrains on the orbit support the necessity to carry out a proper GRAVITY
monitoring program over, at least,  the following 5 years, which
together with the previous 2012 and 2016 data, would allow us to
obtain more accurate estimates of the orbital motion of the
system. The current Gaia TGAS \citep{Gaia1_2016, Gaia2_2016} lists for HD\,93~206 a highly uncertain
parallax of $\epsilon$ = 1.76 $\pm$ 0.62 mas, which includes, e.g.,
the distance to the Carina nebula within its 2$\sigma$
uncertainty. Once more a precise Gaia parallax measurement become
available, it (combined with our new astrometric data) will further improve the constraints on the mass of the stellar components.

The general spectral
signature of HD~93\,206~A obtained with GRAVITY is consistent with the
one taken with FIRE. In both spectra the Br$\gamma$ (2.1661 $\mu$m) shock tracer and HeI (2.058 $\mu$m and 2.112
$\mu$m) photospheric lines
can be identified. The observed line profiles resemble the ones of late O- and early B-stars \citep[see
e.g.,][]{Hanson_1996, Ghez_2003, Tanner_2005b}. This result confirms
the presence of O9.7~I and O8~III stars in
HD~93\,206~A.  However, some differences between
the FIRE and GRAVITY spectra are observed. The HeI line at 2.112~$\mu$m in the FIRE spectrum appears to be an absorption
feature with a well-defined peak and a wing shifted towards the red
part of the spectrum. In contrast, the GRAVITY spectrum shows
an absorption  "W-shaped'' profile. One of the most plausible causes
of these line changes is the difference of the orbital
phase in the compact binaries between both spectra  \citep[see e.g.,
Figure\,1 and Figure\,2 in ][]{Taylor_2011}. Variability of the
spectral features in the visible part of the spectrum of HD\,93~206~A was also reported by
\citet{Morrison_1979} and \citet{Mayer_2001}, associating it to
the double tight-binary nature of the source. 

\begin{figure}
\centering
\includegraphics[width=\columnwidth]{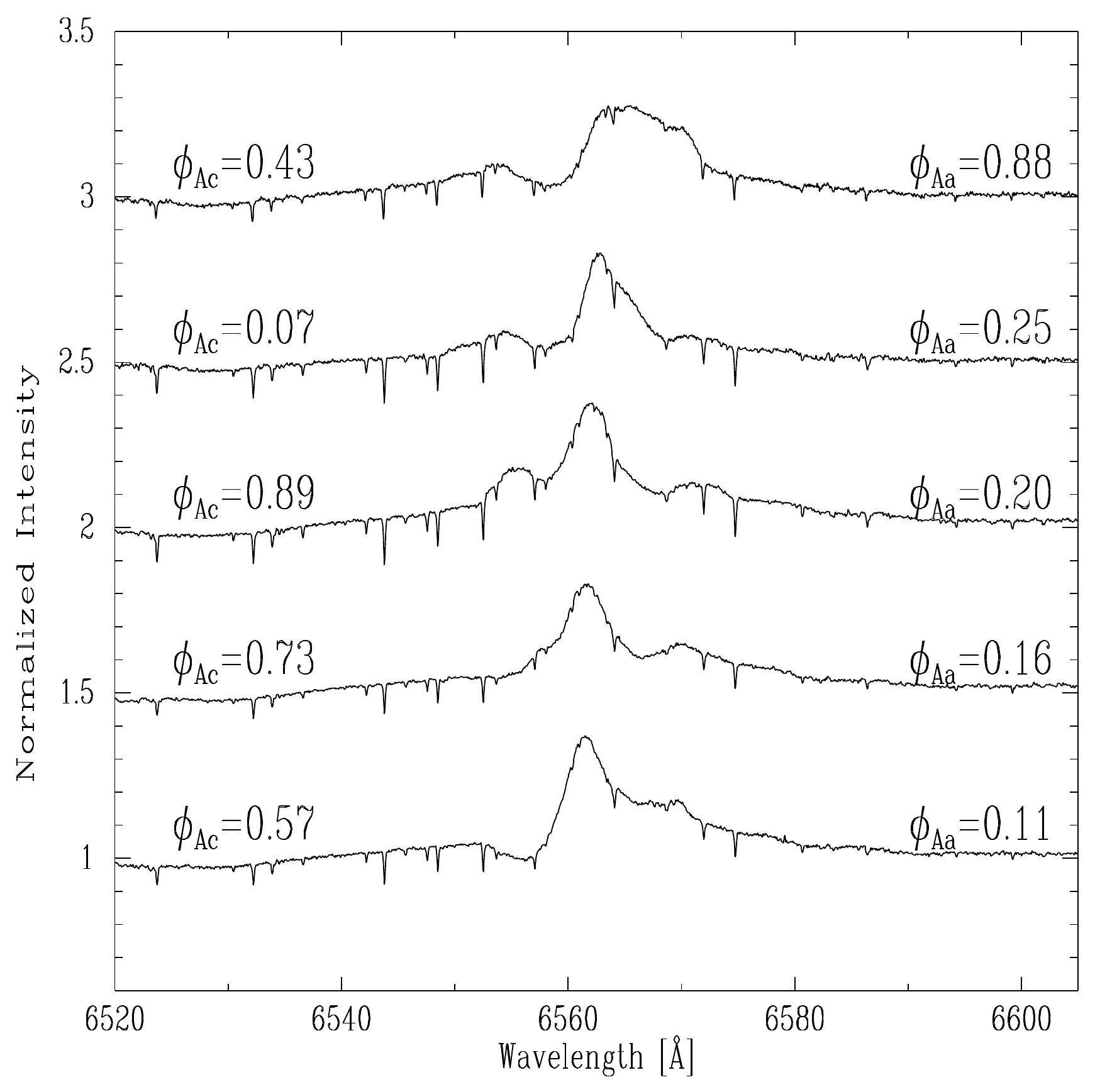}
\caption{H$\alpha$ spectra of HD~93\,206~A obtained with FEROS. The figure displays 5
  spectra at different orbital phases of the two spectroscopic
  binaries that compose HD~93\,206~A. The morphology of the H$\alpha$
  line clearly varies depending on the orbital phase. On the plot, each one of the
  epochs has been shifted vertically for a better representation.}
\label{fig:spec_halpha}
\end{figure}

To test this scenario, Figure\,\ref{fig:spec_halpha} shows normalized
spectra in the H$\alpha$ line (0.656 $\mu$m) of HD~93\,206~A obtained with FEROS (see
Sec.\,\ref{sec:fire_obs}). Considering the ephemeris
  published by \citet{Mayer_2001} for Aa+Ac, we determine the orbital
  phases corresponding to each one of the
spectroscopic binaries. The different phases are labeled to the left and right of each spectrum,
respectively. At first glance, the variations of
the profile on a day to day basis are very noticeable. The 2009 series of spectra (the four spectra from bottom
to the top in Fig.\,\ref{fig:spec_halpha}) were obtained after the
periastron passage of component Aa, but at more random orbital phases
for component Ac. However, for the 2012 spectrum (the top data set in
Fig.\,\ref{fig:spec_halpha}) shows Aa in an orbital phase just before  the
periastron and Ac close to the apastron. From a direct inspection of the overall shape and
changes of the profile, we notice that while the Aa component is in
an orbital phase close to the periastron, the line-peak profile
remains similar in shape and amplitude. However, once the Aa
component is at $\phi_{Aa}$ = 0.88, the line-peak profile becomes
wider. From those observational changes, we infer that most of the narrow
emission, close to the line peak, comes from the Aa component.

On the
other hand, changes in the blue-shifted wing of H$\alpha$ appear to be
related to component Ac. Notice how an absorption is observed in the
wing at orbital phases close to the apastron ($\phi_{Ac}$ = 0.57 and
$\phi_{Ac}$ = 0.43) and how it changes to a secondary emission peak
before the periastron ($\phi_{Ac}$ = 0.89). However, besides the described changes in the line, without a detailed study of both
binary components, it is still challenging to  fully constrain which component of the
binaries is contributing to the emission and/or absorption part of
the profile. Future monitoring of the spectrum not only with GRAVITY
but with other facilities like FEROS or UVES/VLT are necessary to 
complete the spectral analysis of the source.

Additionally to the observed changes in the FEROS data, it is important to
highlight that the Br$\gamma$ line profile
seen with GRAVITY is quite similar to the one of the H$\alpha$ line
observed by
\citet{Mayer_2001} with the 1.4 m CAT telescope at La Silla
Observatory and the bottom spectrum presented in Figure\,\ref{fig:spec_halpha}. Both the Br$\gamma$ and the H$\alpha$ lines have a peak
emission of about 10-20\% above the continuum with a sharp edge at the
blue part of the spectrum and a extended wing towards the
red. The Br$\gamma$ red wing
extends up to $\sim$540 km/s, which is similar to the velocity of the
H$\alpha$ red
wing reported by \citet{Mayer_2001}. This line feature is a very
strong indication for the presence of a stellar wind component.

The X-ray luminosity measured by \citet{Parkin_2011} supports the
presence of shock-tracers like Br$\gamma$ in the wind-wind collision
regions of the quadruple system. However, the semi-detached nature of the
binary Ac suggests the existence of a substantial case-B mass transfer \citep{Leung_1979} in which
the transferred shock-heated plasma might be the main contributor to the observed Br$\gamma$
emission. A possibility to recognize which of the two binaries is the
source of the spectral features is to disentangle the individual
spectra of the binaries Aa and Ac. With
the interferometric observations this is possible following the
procedure detailed in \citet{Chesneau_2014}. This method
  relies, first, on the determination of the binary separation through
a model fitting to all the data (like the one we applied in
Sec.\,\ref{subsec:model_fitting}). Second, once the separation is
determined, one can analyze the relative fluxes C$_i$ of each one of the
components per independent channel. Once C$_i$ have been obtained, one
can get the spectrum S$_i$ of each component using the following
expression:

\begin{equation}
S_i = \frac{C_iS_*}{\sum_{j=1}^{n} C_j}\,
\end{equation}

where S$_*$ is the spectrum of the source. After trying this method, we concluded that the data quality is not good
enough to perform such analysis with the current data sets. Two main
reasons are identified for this limitation: (i) the relatively large variation of
the observables in adjacent channels across the bandpass and (ii) the
relative faintness of the detected lines. However, future observations
with GRAVITY at the UTs would allow us to perform this
analysis. 

\begin{figure*}
\centering
\includegraphics[width=\linewidth]{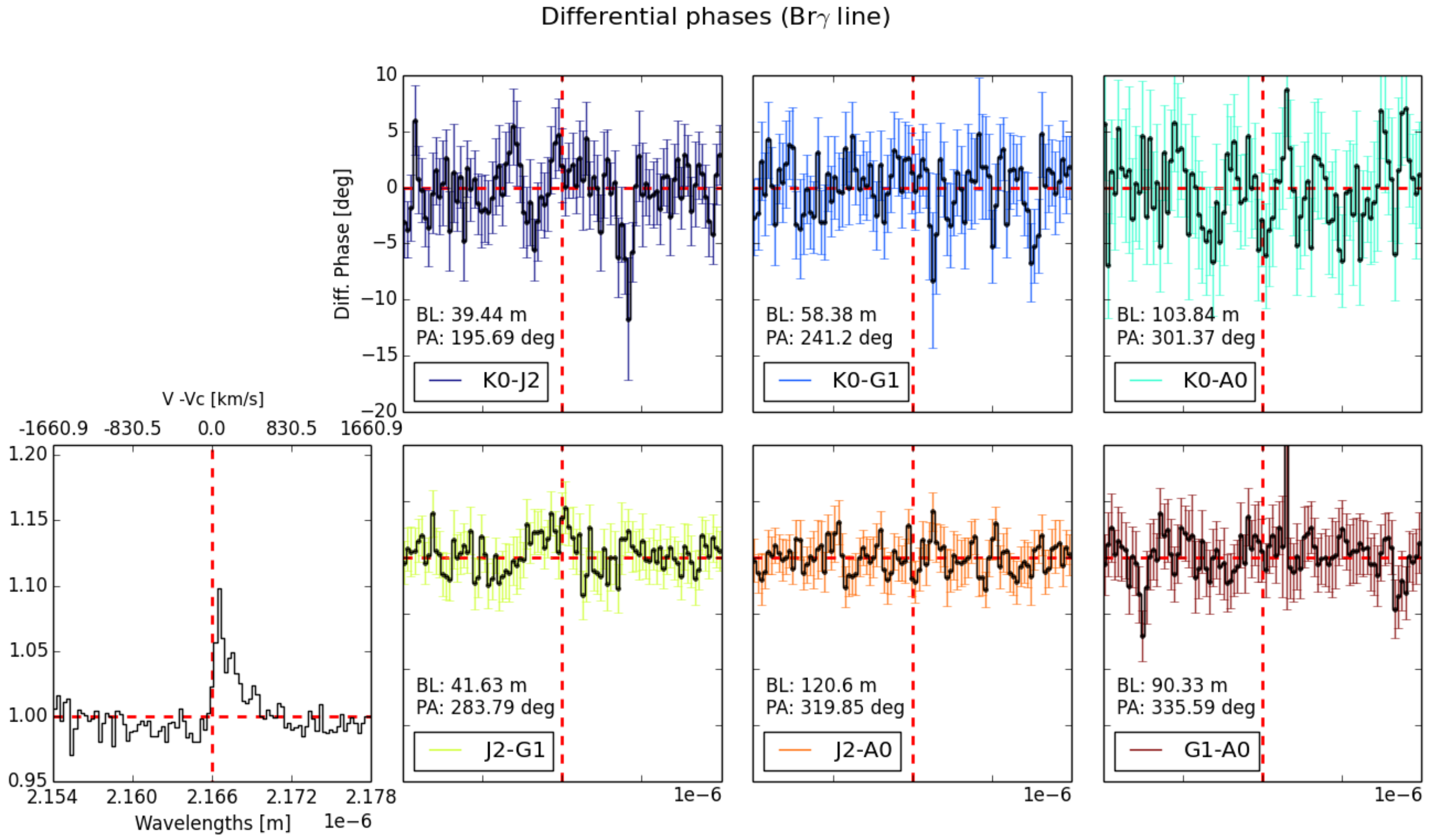}
\caption{The figure displays the differential phases per baseline at
  the position of the Br$\gamma$ line for one of our GRAVITY data sets
(MJD: 57557.046). The left most panel shows the normalized total spectrum of the
source. The upper axis displays the relative velocity from the nominal
position of the line. The continuum baseline is plotted with a
red dashed line. The differential phases are plotted in different
colors depending on the baseline (see labels on the figure). The
baseline length and its position
angle are labeled at each panel. The nominal position of the
Br$\gamma$ line is shown with a vertical red dashed line. The
zero-point in the differential phases is displayed at each panel with
an horizontal dashed line. All the panels share the same scale in the
horizontal axis.}
\label{fig:diff_phases}
\end{figure*}

To investigate the region in which the observed Br$\gamma$ line is
formed, we computed the differential phases
\citep{Millour_2006}. Since we observe the target as a binary, the
pseudo-continuum has a contribution from both components of the
binary; thus, detecting a differential signature implies a relative change in
the brightness of the components at the position of the line. 
Nevertheless, we do not detect any systematic
change in this observable at the position of Br$\gamma$ compared to
the pseudo-stellar continuum. As an example, Figure \ref{fig:diff_phases}
displays the Br$\gamma$ differential
phases of one of our GRAVITY data sets (similar trends are
observed in the other ones). Since the  flux ratio of the
binaries appears to be preserved across the Br$\gamma$ profile, this line is produced at each of the internal binaries instead of being formed in
a more extended region as a consequence of the interaction between both
sub-systems Aa and Ac. However, this result is constrained by the SNR in our data. Considering an average 3$\sigma$ deviation
of the differential phases of $\Delta_{\phi}^{3\sigma}\sim$15$^{\circ}$, we derived an upper limit for the
line-emitting regions of $p$=0.157 mas for a baseline of $B$=120 m,
following the next expression \citep{Kraus_2012_SPIE}:  

\begin{equation}
p = \frac{\Delta_{\phi}^{3\sigma}}{2\pi}.\frac{\lambda}{B}\,.
\end{equation}

Since the observed lines are
highly variable depending on the different orbital phases of the spectroscopic
binaries, future observations of this source with higher
SNR are required to systematically determine the astrometric offsets
associated with the different orbital phases of the inner binaries. With at least
3.5 mag more in sensitivity, GRAVITY using the 8-meter Unit Telescopes
will provide us the required capabilities to (i) monitor the temporal
changes in the lines, (ii) to disentangle the spectra of the
spectroscopic binaries and (iii) to provide an estimation of the
orbital solution for the internal binaries (with a expected
differential phase calibration of
$\Delta_{\phi}^{3\sigma}\sim$2$^{\circ}$ or 20 $\mu$as at 120 m baselines).

\section{Conclusions}
 \label{sec:conclusions}

This work demonstrates the enormous scientific capabilities of the new
GRAVITY instrument fo the characterization of multiplicity in massive
systems. The system HD~93\,206~A is a showcase for this kind of
analysis. It
  represents the first object of a future survey with GRAVITY. In particular, we have obtained the following
results for HD~93\,206~A:  

\begin{itemize}

\item We characterize the relative astrometry between the spectroscopic
  binaries Aa and Ac with a precision of tens of microarcseconds. This corresponds
  to a dramatic improvement of one and three orders of magnitude
  compared to the
  previous astrometric measurements with PIONIER/VLTI and NACO/SAM,
  respectively. The derived separation between Aa and Ac is $\rho \sim$30 mas with
  a position angle PA$\sim$328$^{\circ}$ and a $\Delta
  K$=0.32 mag, with a detected projected motion towards the North-West
  from the 2012 data. This demonstrates the gravitational boundedness
  between the two SBs and provides a rough estimate of their mutual
  orbital period of P $\leq$ 60 years. The precise determination of
  the Aa-Ac astrometry over
  several future epochs ($\sim$5 years) will be crucial to constrain the fundamental
  parameters of the system, its formation history and to predict whether
  the system is long-term (over a few million years) stable or not. 

\item From the \texttt{CANDID} analysis of our GRAVITY observations, we derived a maximum magnitude difference of
  $\Delta K\sim$5 mag to detect companions, with a 3$\sigma$ confidence
  level, in a range of 3 to 50 mas around the primary. This
  represents an increment in sensitivity of $\sim$2 mag from the one
  achieved with PIONIER/VLTI. The current surveys with the 1st
  generation of VLTI instruments have a limiting $\Delta_m \sim$ 3 - 4
  mag between binary components with separations from 1 to 10 mas. Thus, the present result
  confirms GRAVITY as a unique instrument to explore a new parameter space in the
  characterization of massive multiple systems. Here, we have
  demonstrated the superb calibration and sensitivity of the
  single-field mode of the instrument. Nevertheless, the expected
  improvement in the GRAVITY performance (particularly its complete
  integration with the 8 m Unit Telescopes) over the next year,
  in combination with the so-called ``dual-field'' mode of the
  instrument, will provide us the opportunity to look for companions
  in stars as faint as $\sim$10 mag with the ATs and $\sim$16 mag with the UTs, letting us to
  completely characterize systems like HD\,93\,206.
 
\item A dedicated Principal Component Analysis of the GRAVITY data
  allowed us to extract the calibrated spectrum of the source in which
  shock tracers (Br$\gamma$) and stellar (HeI) lines are observed. The
  detected spectral features have a maximum intensity $\sim$10\%
  above the pseudo-stellar continuum. Complementary FIRE data supports
  variability of the $K-$band lines and visible FEROS spectra indicate
  that the daily variability is linked to the motion of the
  multiple components of the quadruple. It appears that changes around the peak of the H$\alpha$
  are associated to the Aa component, while changes in the wings
  correspond to the Ac component.

\item We constrain the near-infrared wind-wind collision region
  associated to the observed X-ray emission of the source. From our differential phase analysis at the position
  of Br$\gamma$, we confirm that it is formed in very compact
  regions inside the inner binaries and not from a more extended
  region between both sub-systems Aa and Ac. This is consistent with
  previous models which suggest that the X-ray emission is arising
  mostly within the spectroscopic binaries, in particular from
  component Ac. With the current data we could not confirm which of
  the two spectroscopic binaries is the main X-ray emitter. However, from the 3$\sigma$ standard deviation of the
  differential phases from the continuum baseline, we infer an upper
  limit for the line-emitting regions of p$\sim$0.157 mas (0.37 AU). 

Future observations with the UTs in ``single-field'' mode will
  provide us a calibration of $\Delta \phi
  \sim$1$^{\circ}$ to constrain the line-emitting
  region up to scales of 20 $\mu$as (0.05 AU at the distance of the source). If the detected changes in H$\alpha$ are
  applicable to Br$\gamma$, detecting systematic
  changes in the differential
  phases would allow us to disentangle the astrometric motion of the
  individual components within the spectroscopic binaries. Monitoring
  projected astrometric motions combined with spectroscopic data and a
  proper model of the emission line variability will let us
  to characterize, for the first time, the fundamental parameters of
  the individual components in the system. This information would
  provide the mass distribution among the components, test coplanarity
  among the different orbital planes and confront the different formation
  scenarios in this kind of systems. 

\end{itemize}




\acknowledgments
We thank the anonymous referee for his/her useful comments to improve
our manuscript. We thank the ESO and GRAVITY consortium Science Verification Team
 for all their support to carry out the presented GRAVITY
 observations. We thank to Gabriel Ferrero for his help in processing
 the FIRE spectra. This research has made use of the Jean-Marie
 Mariotti Center \texttt{SearchCal} and \texttt{LitPro} services co-developed by FIZEAU and
 LAOG/IPAG, CRAL and LAGRANGE. J.S.B acknowledges the support from the Alexander
 von Humboldt Foundation Fellowship programme (Grant number ESP
 1188300 HFST-P); this work was also
partly supported by OPTICON, which is sponsored by
the European Commission's FP7 Capacities programme
(Grant number 312430).  J.M.A. acknowledges support from the Spanish Government Ministerio de Econom{\'\i}a y Competitividad (MINECO) 
through grant AYA2013-40\,611-P. A.A. acknowledges support from the Spanish Government Ministerio de Econom{\'\i}a y Competitividad (MINECO) 
through grant AYA2015-63939-C2-1-P, co-funded with FEDER funds. 
F.C. acknowledges that the research leading to these results has
received funding from the European Community's Seventh Framework
Programme under Grant Agreement 312430 (OPTICON). The research leading
to these results has received funding from the European Research
Council under the European Union's Seventh Framework Programme
(FP7/2007-2013) / ERC grant agreement No. [614922].

\software{GRAVITY pipeline \citep{Lapeyrere_2014}, esorex
  (http://www.eso.org/sci/software/cpl/\\esorex.html), P2VM algorithm \citep{AMBER_Tatulli_2007, Lacour_2008}, FIREHOSE, MIDAS, emcee \citep{Foreman_2013}, Scipy (http://www.scipy.org/), LitPro \citep{LITpro}, CANDID \citep{Gallenne_2015}}

\bibliographystyle{aasjournal}
\bibliography{/Users/bluedemon/Documents/Papers/Paper_lib}

\appendix
\section{SNR distributions of the interferometric observables}
\label{sec:SNR_obs}

\begin{figure*}[hp]
\centering
\includegraphics[width=13.5 cm]{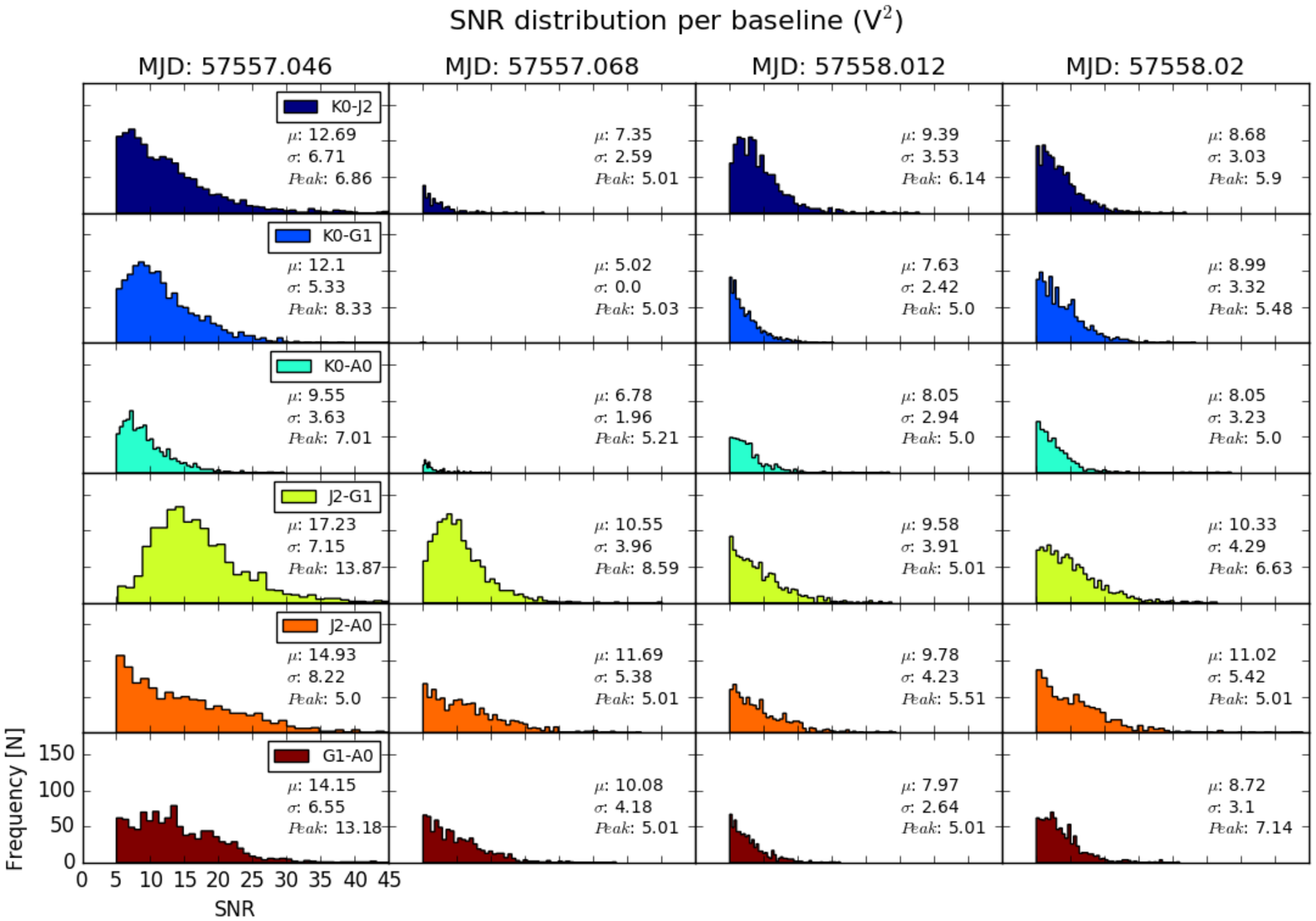}
\caption{SNR distributions of the V$^2$. Every column in the plot
  corresponds to a different GRAVITY data set. Histograms for
  different baselines are plotted in different colors (see label on
  the figure). The values of the mean and standard deviation of the SNR, as well as
  of the peaks of the distributions are printed on each panel in the figure.}
\label{fig:snr_v2}
\end{figure*}

\begin{figure*}[hp]
\centering
\includegraphics[width=13.5 cm]{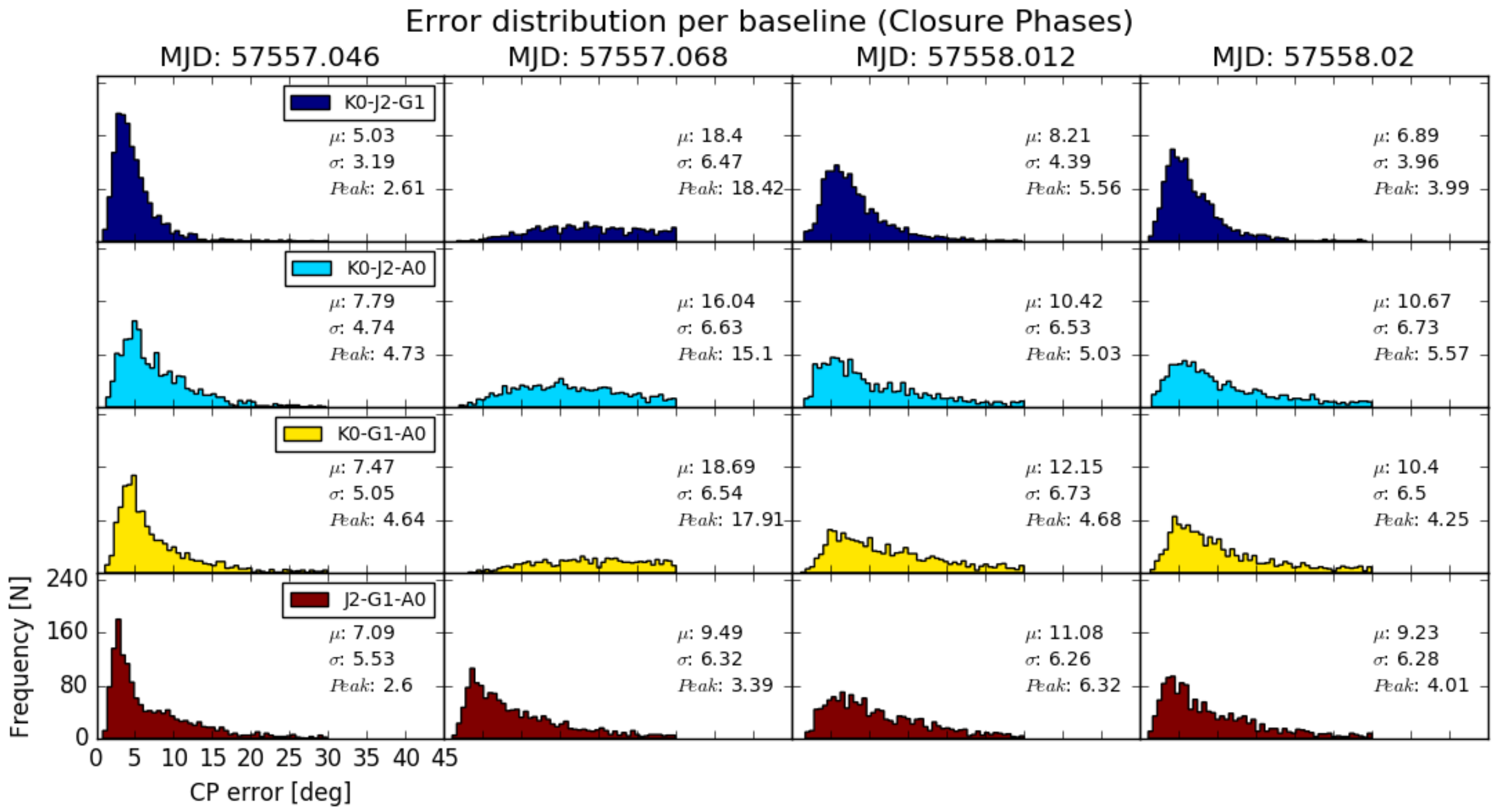}
\caption{Closure phase error distributions. Every column in the plot
  corresponds to a different GRAVITY data set. Histograms for
  different baselines are plotted in different colors (see label on
  the figure). The values of the mean and standard deviation of the
  closure phase errors, as well as of
  the peaks of the distributions are printed on each panel in the figure.}
\label{fig:snr_cps} 
\end{figure*}

\end{document}